\documentclass{hep99}
\usepackage{epsfig}
\def\nostrocostrutto#1\over#2{\mathrel{\mathop{\kern 0pt \rlap%
  {\raise.2ex\hbox{$#1$}}}
  \lower.9ex\hbox{\kern-.190em $#2$}}}
\def\gsim{\nostrocostrutto > \over \sim}   
\begin{document}
\title{Lightest glueball and scalar meson nonet
in production and decay\footnote{
presented at EPS-HEP '99 Conference, July 15-21, 1999, Tampere, Finland,
MPI-PhT/99-40} 
}
\author{Wolfgang Ochs}
\address{Max Planck Institut f\"ur Physik,  Werner Heisenberg Institut,
    D-80805 Munich, Germany
\\[3pt]
E-mail: {\tt wwo@mppmu.mpg.de}}
\abstract{
Recent results concerning the evidence and  classification of the 
$J^{PC}=0^{++}$ states, obtained with P. Minkowski, are presented:
The isoscalars $f_0(980)$ and $f_0(1500)$ are classified
as members of the $0^{++}$ nonet,
while the broad state
called $f_0(400-1200)$  and the state $f_0(1370)$ are
considered as different components of a single broad resonance,
the lowest-lying $0^{++}$ glueball. Furthermore, we propose
the investigation of glueball production
in the fragmentation region of gluon jets.
}
\maketitle

\section{Introduction}
A longstanding problem in hadron spectroscopy is the existence of gluonic
mesons or ``glueballs''. Despite many experimental efforts
not a single such state is generally accepted in the community until now.

On the theoretical side the existence of such states  is
considered  a necessary consequence of the QCD dynamics \cite{HFPM} and in most
considerations it is 
concluded that the state of lowest mass should have the quantum
numbers $J^{PC}=0^{++}$.

Some convergence of the numerical results from lattice QCD has been
reached in recent years in the application of the 
quenched approximation, i.e. neglecting fermion loops,
with mass values for the $0^{++}$ glueball in the range 1600-1700 MeV 
(for a recent review, see Teper \cite{Teper}). 
This prediction has also influenced the experimental searches
in the last years. On the other hand, recent lattice calculations which
include two light quark flavors \cite{lattunq} suggest the glueball
mass to decrease with the
quark mass; the ultimate conclusion from this approach
has therefore to await the final results from the unquenched calculations.

An alternative approach to the properties of hadrons are the QCD
sum rules \cite{svz}. In a recent application by Narison \cite{Nar} 
it was found impossible to saturate the relevant sum rules with a
single $0^{++}$ glueball at a mass around 1500 MeV alone. Rather, the
inclusion of a light gluonic component in the mass region around 1 GeV was
required. The role of a light glueball in the saturation of the QCD
sum rules was already pointed out some years ago by Bagan and 
Steele \cite{bagan}.  

In view of this situation we have reconsidered the spectroscopy of the light 
$0^{++}$ hadrons and we do not exclude the states below 1500 MeV as candidates
for the lightest glueball or for 
a state with large gluonic admixture \cite{mo}.
    
\section{Evidence for light scalar isoscalar states}
The Particle Data Group (PDG \cite{PDG}) lists the following states
below 1600 MeV mass: $f_0(400-1200),\ f_0(980),\ f_0(1370)$ and
$f_0(1500)$. To establish the resonance it is important to observe
the characteristic variation of the scattering amplitude: it moves locally
around a circle in the complex plane. In elastic and inelastic 2-body
scattering additional constraints are provided by unitarity. Therefore, we
have concentated on the appearence of the above states in $\pi\pi$
scattering processes into final states $\pi\pi,\ K\overline K,\ \eta\eta$
which can be obtained from the analysis of peripheral pion exchange
processes.

There is no controversy about the existence of the narrow $f_0(980)$
which appears as a dip in the elastic $\pi\pi$ cross section.
The construction of the scattering amplitudes for
 $\pi\pi\ \to \ K\overline K,\ \eta\eta$ in the complex plane (``Argand
diagram'') reveals in both channels a 
Breit-Wigner resonance centered
around 1500 MeV above a slowly moving background with negative ($ K\overline
K$) or positive sign ($\eta\eta$) relative to $f_2(1270)$, see Fig. 1.
These amplitudes correspond to mass spectra which produce peaks around
1300-1400 MeV in the $K\overline K$ and 1600 MeV in the $\eta\eta$ channels.
Only the state $f_0(1500)$ associated with the full circle is considered 
 a genuine
resonance; the other structures, especially the $f_0(1370)$ we take
-- together with the $f_0(400-1200)$ -- as
components of a single  broad state which extends from 400 up to 1500
MeV or beyond and only in this large mass interval 
the elastic $\pi\pi$  amplitude describes a full circle
as required for a Breit Wigner resonance. The mass and width of this 
state is estimated as 
\begin{equation}
m\sim 1000\ {\rm MeV},\quad \Gamma\sim 1000\ {\rm MeV} \label{gb1000}
\end{equation}

\begin{figure}[t]
\begin{center}
\mbox{\epsfig{file=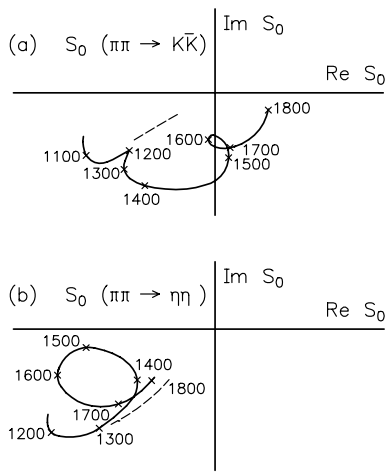,width=50mm}}
\end{center}
\caption{\it
Argand diagrams of the isoscalar $S$-wave amplitudes
constructed \protect\cite{mo}  
from data on the mass spectra 
and the relative phases between $S$ and $D$-waves \protect\cite{bnl}
assuming a Breit-Wigner
form for the latter.
The numbers indicate the pair masses in MeV, the dashed curves indicate an
estimate of the background.   
\label{gbfig4} }
\end{figure}

\section{The $0^{++}$ $q\overline q$ nonet}

As members of the nonet we take the isoscalars $f_0(980)$ and $f_0(1500)$,
in addition the isovector $a_0(980)$ and the strange $K^*(1430)$. The mixing
pattern is close to the one found in the pseudoscalar sector with
flavour amplitudes $(u\overline u,d\overline d,s\overline s)$:
\begin{equation}          
  \label{mixing}           
  \begin{array}{llll}      
  f_0(980) &\leftrightarrow & \eta^{\prime} (958)  & \sim \ 
\frac{1}{\sqrt{6}}(1,\ 1,\ 2) 
      \\
  f_0(1500)& \leftrightarrow & \eta (547)  & \sim \ 
\frac{1}{\sqrt{3}}(1,\ 1,\ -1) 
  \end{array} 
\end{equation}
i.e. the $f_0(980)$ is near the singlet, the $f_0(1500)$ is near the octet
state.
This classification has been suggested and is supported by the following
observations:

\noindent {\it 1. $J/\psi\to \omega,\varphi +X$ decays}\\
The branching ratios of  $J/\psi$ into $\varphi \ \eta^{ \prime}  (958)$ and
$\varphi \ f_{ 0}  (980)$ are of similar size and about twice as large as
$\omega \ \eta^{ \prime}  (958)$ and  $\omega \ f_{ 0}  (980)$. These
observations are reproduced by the flavor composition (\ref{mixing}).

\noindent{\it 2. Two body decays of $f_0$ states}\\
Given the flavor composition (\ref{mixing}) we can derive the
decay amplitudes  for $f_0(980)\to \pi\pi,\ K\overline K$, as well as for
 $f_0(1500)\to \pi\pi,\ K\overline K,\ \eta\eta,\ \eta\eta'$. The 
available data are
reasonably well reproduced if  we allow
for a $s\overline s$ relative amplitude in the range $S=0.5\ldots 1.0$.
The decays  $f_0(980)\to \gamma\gamma$ and   $a_0(980)\to \gamma\gamma$
are important but not yet very decisive.

\noindent {\it 3. Relative signs of decay amplitudes}\\
A striking prediction is the relative sign of the decay amplitudes of the
$f_0(1500)$ into pairs of pseudoscalars: because of the negative sign in the
$s\overline s$ component, see eq.(\ref{mixing}), the sign of the
$K\overline K$  decay amplitude is negative with respect to $\eta\eta$
decay and                  
also to the respective $f_2(1270)$ and glueball decay amplitudes.
This prediction is indeed confirmed by the amplitudes in Fig.1 which show
 circles pointing in upward and downward directions, respectively.
If $f_0(1500)$ were a glueball, then both circles should
have the same sign, but the experimental results are rather
orthogonal to such an expectation.

\noindent {\it 2. Gell-Mann-Okubo mass formula}\\
This formula is satisfied for the squared
masses of our flavour octet built from
$a_0(980),K^*(1430)$ and $f_0(1500)$ within $\sim$10\%.

The mass pattern inside this scalar nonet is found consistent with
what is obtained from a general effective QCD potential, not restricted to
renormalizable interactions, after expansion to first order in the strange
quark mass \cite{mo}. 
We also note that the near octet flavor mixing of the
$f_0(1500)$ and its heavy mass can be
understood in a 3-flavor Nambu-Jona-Lasinio model with an instanton induced
axial $U(1)$ symmetry breaking term \cite{njl}.

\section{The lightest $0^{++}$ glueball}
The remaining states in the PDG tables, the $f_0(400-1200)$ and $f_0(1370)$,
are not accepted as distinct Breit-Wigner resonances because of a lack of
sufficient phase variation of the amplitude. Rather, they are considered as
two components of a yet broader state with parameters (\ref{gb1000})
which also has an inelastic coupling, visible as slowly moving
``background'' in the reactions shown in Fig.1. It is our
hypothesis that this broad state is the lightest glueball
$gb(1000)$. A mixing with the
nonet states is not excluded but it is expected to be sufficiently small
such that the structures outlined before are not destroyed. The following
observations support the identification of this state with the glueball:

\noindent{\it 1. Reactions favorable for glueball production}\\
The broad object $gb(1000)$ is observed in the ``gluon rich'' processes:
the central production in hadron-hadron collisions which are expected to
proceed through double-Pomeron exchange; in the decays of radially excited
Onia $\psi'\to\psi(\pi\pi)_s$ and $Y',Y''\to Y(\pi\pi)_s$ which are 
expected to proceed through gluonic exchanges. 
On the other hand, no prominent signal is observed in the radiative 
$J/\psi$ decays as expected, but the statistics is rather low. 
   
\noindent{\it 2. Suppression in $\gamma\gamma$ collisions}\\
If the mixing of the glueball with charged particles is small it should be
weakly produced in $\gamma\gamma$ collisions. 
In the process $\gamma\gamma\to \pi^0\pi^0$ there is
a dominant peak related to $f_2(1270)$ but,
in comparison to elastic $\pi\pi$ scattering, a very small cross section
in the low mass region around 600 MeV. 
In a fit to the data which takes into account the
one-pion-exchange Born terms and $\pi\pi$ rescattering Boglione and
Pennington \cite{BP} have determined 
the two photon width
of the states $f_2(1270)$ and $f_0(400-1200)$ 
as 2.84$\pm$0.35 and 3.8$\pm$ 1.5 keV, respectively.
If the $f_0$ were a light quark state like the $f_2$ we might expect
comparable ratios of 
$\gamma\gamma$ and $\pi\pi$ decay widths
$R(f)=\Gamma(f\to\gamma\gamma)/\Gamma(f\to\pi\pi)$,
but we find
\begin{equation}           
R(f_2)
  \sim \  15\times 10^{-6} ;\quad        
R(f_0)
\sim \ 4 \ldots 6 \times 10^{-6}, 
\label{R02}                
\end{equation}             
thus, for the scalar state, this ratio is 3-4 times smaller, and it could be
smaller by another factor 3 at about the 2$\sigma$ level. So, the two photon
coupling of the $gb(1000)$ is indeed smaller than one would anticipate for a
$q\overline q$ state.
 
Further support for the glueball hypothesis is drawn \cite{mo} from the
large width of $gb(1000)$, certain branching ratios of ``$f_0(1370)$'' and
the non-$q\overline q$ component in elastic $\pi\pi$ scattering.

\section{Gluon fragmentation into glueballs}
The observations reported above and others in \cite{mo} are 
largely consistent with
the proposed identification of the scalar states.
We suggest studying further the quark and gluon constituent nature of these
states. An attractive environment not yet explored
is the fragmentation region of gluon and
quark jets, respectively, with resonances produced at large energy fractions,
 say with
\begin{equation}
z=E_{resonance}/E_{jet},\qquad z\gsim 0.5.  \label{frag}
\end{equation}
The production of color singlet clusters in gluon jets has been
considered already in \cite{pw} but the idea applied to glueballs
has not been pursued further.
 
In the mass spectra of particles at large 
total $z$ the resonances should show up
clearly with little combinatorial
background; of particular interest for the glueball search
are the
$\pi\pi$ and $K\overline K$ pairs, especially in the neutral mode
to select $0^{++}$ and $2^{++}$ states.

It is well established that a quark fragments at large $z$ predominantly 
into a
hadron with the primary quark as valence quark. Similarly, a gluon 
at large $z$ might fragment into a glueball with the primary gluon as
valence gluon. Then it will be interesting to compare resonance production
in quark and gluon jets at large $z$:
a gluonic state like $gb(1000)$ or any higher mass glueball state
should be produced with larger rate in the gluon jet at large $z$ than
in the quark jet and vice versa for $q\overline q$ states like $f_0(980)$
or $f_0(1500)$.

A large number of gluon and quark jets should be available for such 
studies in $e^+e^-\to 3$ jets and in high $p_T$  production of quark and gluon
jets at $pp$ or $ep$ colliders.

\section{Conclusions}
Our analysis of the low lying scalar isoscalar particles \cite{mo}
suggests including  $f_0(980)$ and  $f_0(1500)$ in the $q\overline q$ nonet
with mixing similar to the pseudoscalars $\eta'$
(near singlet) and $\eta$ (near octet), respectively. We concluded the 
$f_0(1370)$ not to be a genuine Breit-Wigner resonance and 
the $ f_0(1500)$ not to
be a glueball or anything close to it.

The states in the PDG called $f_0(400-1200)$ and $f_0(1370)$ are components
of a single broad Breit-Wigner resonance which is a
respectable candidate for the lightest glueball: $gb(1000)$.
The basic triplet of
light binary glueballs is completed in our approach 
\cite{mo} by the states $\eta(1440)$ with
$0^{-+}$ and $f_J(1710)$ with $2^{++}$, not discussed here.

A promising field of further glueball studies is in 
the fragmentation region of
quark and gluon jets.

\end{document}